\documentclass[twocolumn,english]{revtex4-2}
\usepackage[T1]{fontenc}
\usepackage[latin9]{inputenc}
\usepackage{float}
\usepackage{amsmath}
\usepackage{graphicx}

\usepackage{babel}
\begin{document}
\title{Training nonlinear elastic functions: nonmonotonic, sequence dependent
and bifurcating }
\author{Daniel Hexner}
\affiliation{Faculty of Mechanical Engineering, Technion, 320000 Haifa, Israel}
\begin{abstract}
The elastic behavior of materials operating in the linear regime is constrained, by definition, to operations that are linear in the imposed deformation. 
Though the nonlinear regime holds promise for new functionality, the design in this regime is challenging. In this paper we demonstrate that a recent approach based on training
[Hexner \textit{et al., PNAS} 2020, 201922847] allows responses that are inherently non-linear. By applying designer strains, a disordered solids evolves through plastic deformations 
that alter its response. We show examples of elaborate nonlinear training paths that lead to the following functions: (1) Frequency conversion (2) Logic gate and (3) Expansion or contraction along
 one axis, depending on the sequence of imposed transverse compressions. We study the convergence rate and find that it depends on the trained function.
\end{abstract}
\maketitle

\section{Introduction}

Recently, it has been realized that nontrivial functions
can be imparted in ordinary disordered materials through small
microscopic alterations\citep{goodrich2015principle,rocks2017designing,hexner2018role,yan2017architecture,meeussen2020topological}.
This is believed to stem from the enormous design space defined by
the accessible microstates. Originally conceived in tuning
global material properties\citep{goodrich2015principle} (e.g., Poisson's ratio) ,
it was then followed by designing spatially varying responses \citep{mitchell2016strain,rocks2017designing,yan2017architecture,rocks2019limits}
motivated by allostery in proteins. Both global and spatial dependent responses have been realized in simulations and experiment\citep{rocks2017designing,reid2018auxetic}.

Designing the microstructure often employs an optimization process
to compute the structure. Realizing these designs, however, is challenging.
It requires a precise knowledge of the interactions\citep{reid2018auxetic}.
Scaling up the number of degrees becomes numerically expensive, while
shrinking down the size of the microscopic elements necessitates control
on tiny length scales. 

To address these issues, an alternative has recently been proposed\citep{pashine2019directed}, termed "directed aging", which is based on the self-organization of microstructure. It considers disordered materials that deform plastically in response to applied strains. By applying carefully selected sequences of strains, material properties can be steered towards a
desired goal. To date, this approach has been useful in manipulating
global material properties in both simulations and experiment \citep{lakes_r_Science_1987,pashine2019directed,hexner2020effect} (primarily the Poisson's ratio), as well as spatially varying allosteric responses in simulations\citep{hexner2019periodic}. 

In this paper we focus on the nonlinear regime, which allows behaviors that are inherently different than those in linear response \citep{bertoldi2010negative,miura1985method,chen2014nonlinear,grima2000auxetic,coulais2016combinatorial,kim2019conformational,bar2020geometric}. We study the feasibility of training distinctly nonlinear responses by defining training sequences that couple source degrees of freedom to  target degrees of freedom. Our aim is to realize a given strain on the target degrees of freedom in response to a strain on the source degrees of freedom. The training sequence defines curves (or training paths) in terms of the source and target strains. By devising elaborate training paths that are  nonlinear, and may bifurcate, we are able to realize highly non-trivial responses: 1. frequency conversion 2. logic gates and 3. a material that either expands or contracts along one axis, depending on the sequences of applied compression. We study the convergence
as a function of the number of training cycles and find that it is
often very slow.

\section{Model and training procedure}
We model an amorphous solid
using a disordered network of springs whose the energy depends on
the stiffness, $k_{i}$, the bond length $\ell_{i}$ and the rest
length $\ell_{i,0}$: 
\begin{equation}
U=\frac{k_{i}}{2}\sum_{i}\left(\ell_{i}-\ell_{i,0}\right)^{2}.
\end{equation}
We employ networks derived from packings, since their coordination number is easily tuned by the pressure exerted on the box\citep{Ohern}. Details of the network and packing preparation are defined in Appendix A. We allow plastic deformations that
alter the structure through changes to the rest lengths\citep{hexner2020effect}.
The rest lengths of a bond is assumed to evolve in proportion to the
stress it experiences: 

\begin{equation}
\partial_{t}\ell_{i}=\gamma k_{i}\left(\ell_{i}-\ell_{i,0}\right).
\end{equation}
A bond elongates under tension and shortens under compression. This
model is essentially the Maxwell model for viscoelasticity\citep{maxwell1867iv}:
each bond is a spring and dashpot in series, where the latter accounts
for the change in rest length. We assume that the time to reach force
balance can be neglected with respect to the viscous time scales. 

For each response a target deformation and source deformation are
defined, each with a corresponding strain $\epsilon_{T}$ and $\epsilon_{S}$.
These degrees of freedom may either be global (e.g., uniaxial strain) or local  (e.g., squeezing a pair of nodes). Motion in the quasistatic regime follows low energy
directions and therefore training aims at programming low energy ``valleys''
that steer the response. We follow the training procedure of Ref.
\citep{hexner2019periodic} where the system is strained periodically
along the prescribed training paths. We note that, Eq. 2 is the gradient
of the energy with respect to the rest lengths, and therefore at a
constant imposed strain minimizes the elastic energy at that strain. Straining periodically
along a path $\epsilon_{T}=f\left(\epsilon_{S}\right)$ reduces the
energy along that entire path, creating an energy valley. Actuating
the source  forces the system along that same curve $\epsilon_{T}=f\left(\epsilon_{S}\right)$. 

\begin{figure}
\includegraphics[scale=0.57]{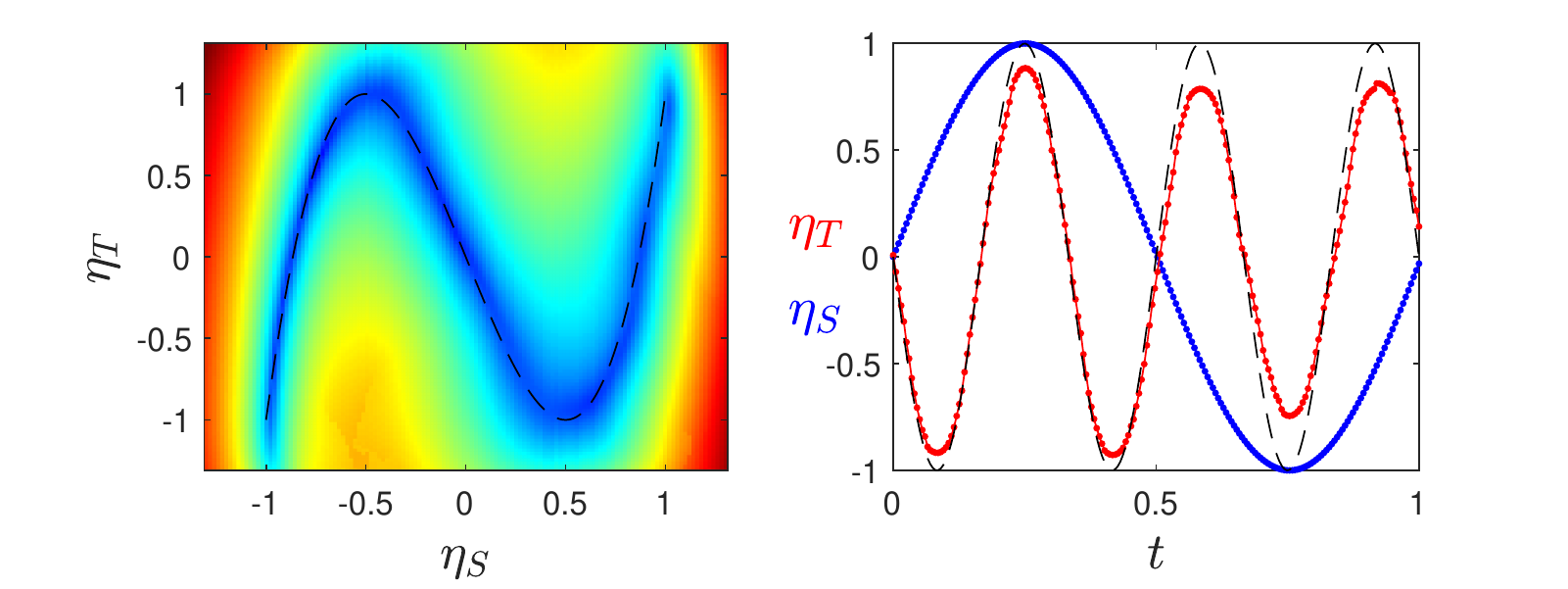}

\caption{Frequency conversion through non-monotonic training path for allostery.
Left: The elastic energy measured as a function of the source and
target strains (normalized to the strain amplitude $\epsilon_{Age}$).
The energy landscape forms a valley along the trained path, $\eta_{T}=-3\eta_{S}+4\eta_{T}^{3}$,
marked by the dashed line. Right: The quasistatic response to an applied
time periodic source strain. The parameters are the same as in Fig.
2. \label{fig:energy} }
\end{figure}
We consider four examples of responses. Of these two are global responses,
and two are spatially dependent that couple far away degrees of freedom.
\section{Results}

\subsection{Frequency conversion through non-monotonic paths}

We begin by training a family of responses whose non-linearity can be tuned,
interpolating between a linear function ($\alpha=0$) and non-monotonic
function ($\alpha>1)$:

\begin{equation}
\eta_{T}=\left(1-\alpha\right)\eta_{S}+\alpha\eta_{S}^{3}.
\end{equation}
For simplicity we have defined $\eta_{S}=\epsilon_{S}/\epsilon_{Age}$
and $\eta_{T}=\epsilon_{T}/\epsilon_{Age}$ where $\epsilon_{Age}$
is the strain amplitude, chosen to be the same on the source and target. 

\begin{figure}
\includegraphics[scale=0.67]{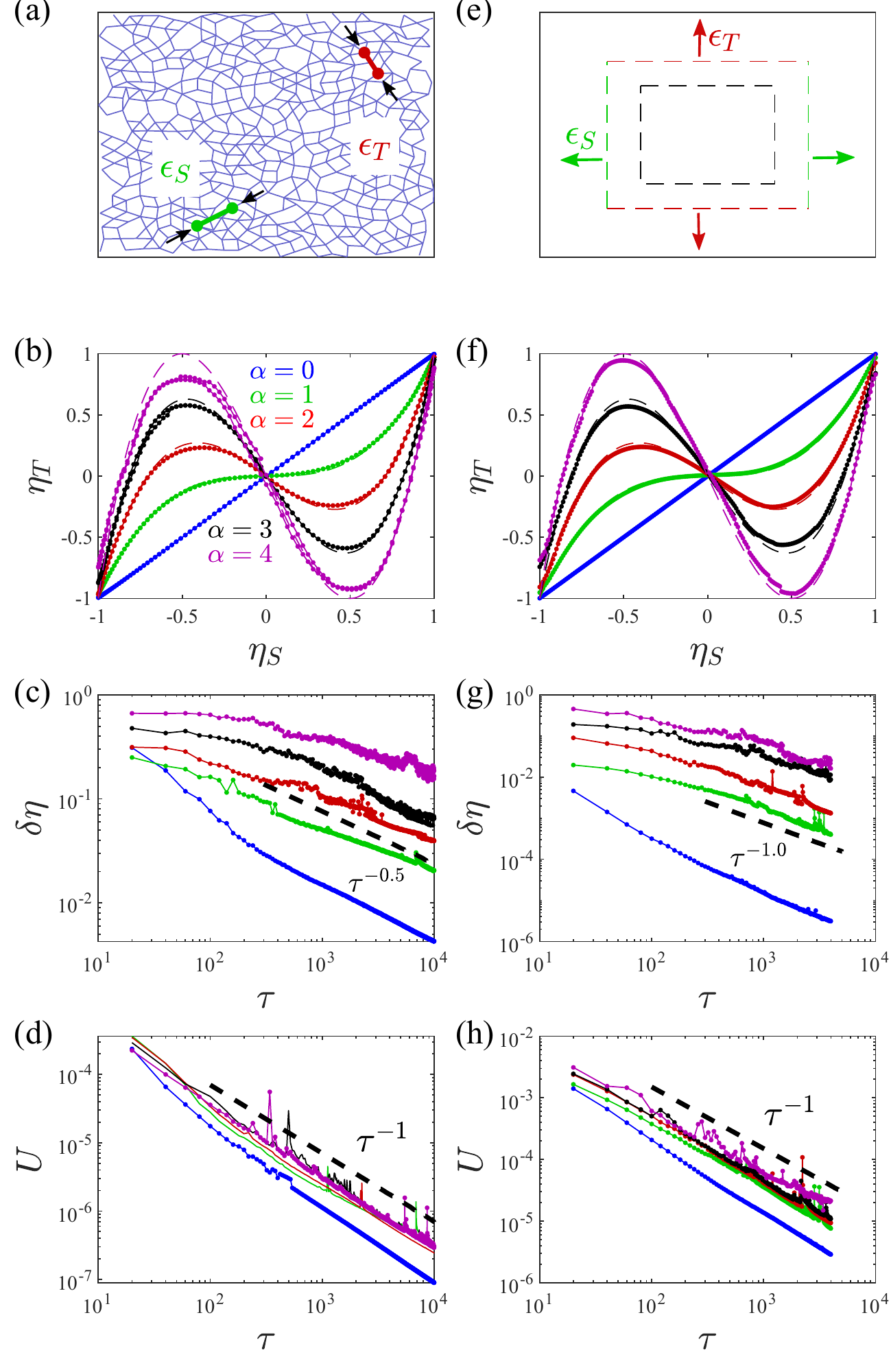}

\caption{Training non-monotonic responses in allostery (left) and uniaxial
response (right). (a) An illustration of the networks and the trained
allostery inspired response. Squeezing the source sites leads to a
prescribed motion in the target nodes. (e) The source and target strain
in the global responses are defined to be the uniaxial global strain.
(b), (f) The training paths are denoted by the dashed lines and the
trained responses with the measured responses overlaid. (c) (g) The
convergence of the trained response as a function of the number of
training cycles. The asymptotic decay appears to be different for
the global and local response (dashed line are a guide to the eye).
(d),(h) shows the decays of the energy as function of the training
cycles. For the global response we choose: $\epsilon_{Age}=0.05$
and the aging rate $\Gamma=16.0$ . For the allosteric response $\epsilon_{Age}=0.2$
and $\Gamma=4.0$. \label{fig:non_mono}}
\end{figure}

To demonstrate the generality of training nonlinear responses, we
consider both the allostery inspired response\citep{rocks2017designing,yan2017architecture,hexner2019periodic}
(Fig. \ref{fig:non_mono} left) as well as global deformations (Fig.
\ref{fig:non_mono} right). The global degrees of freedom are the
uniaxial strain; $\epsilon_{S}$ is the strain along the x-axis while
$\epsilon_{T}$ is the strain along the y-axis. For the allosteric
response the source and target degrees of freedom are pairs of nearby
nodes, as illustrated in Fig \ref{fig:non_mono} (a). Squeezing a
pair of source nodes leads to a prescribed strain on the far away
target nodes. The strain is defined as the fractional change in distance
between a pair of nodes, $\epsilon=\text{\ensuremath{\frac{\ell-\ell_{0}}{\ell_{0}}}}$.
To communicate strain over long distances we are limited to using
near isostatic networks whose anomalous elasticity is long ranged
\citep{ellenbroek2006critical,lerner2014breakdown,hexner2019periodic}.

We focus on $\alpha\leq4$, where $\eta_{T}$ is in the range $\left[-1,1\right]$.
The case of $\alpha=4$ is particularly interesting, since applying
a time dependent sinusoidal source strain $\eta_{S}=sin\left(\omega t\right)$
yields $\eta_{T}=-sin\left(3\omega t\right)$. As noted, during training
both the source and targets are strained periodically which results
in an energy valley that follows the training path. Fig.\ref{fig:energy}(a)
shows this energy landscape, as function of the source and target strains, for the case $\alpha=4$ and allosteric degrees of freedom. To asses the success of training we then actuate
the source degrees of freedom and measure the strain on the target.
Fig.\ref{fig:energy}(b) shows indeed that applying to the source a sinusoidal time dependent strain results in a response on the target at nearly triple the frequency. Other
integer conversion ratios can be achieved by the appropriate polynomials. Thus, we are able to convert the frequency of the applied driving. We note however,  the change in frequency is due to the non-monotonic energy valley -- all our responses are in the quasistatic regime. 

We now turn to discuss the behavior as a function of $\alpha$. The
success of training is characterized in Fig. \ref{fig:non_mono}(b)
and (f) following the training period. We plot $\eta_{T}$ as a function
of the $\eta_{S}$ averaged over approximately 50 of samples. On average
the response is close to the training paths, denoted by the dashed
lines. Training is more more successful for small values of $\alpha$. 

To quantify the convergence of the response as a function of the number of training cycles we define a quantity  $\delta\eta$ that measures the distance of the response from the intended value.  We strain the source degrees of freedom, by
varying the strain in $M$ small strain steps, denoted with $i$ and
measure the deviation of $\eta_{T}$ from the trained path: 
\begin{equation}
\left(\delta\eta\right)^{2}=\frac{1}{M}\sum_{i=1}^{M}\left(\eta_{T}^{\left(i\right)}-\eta_{T,path}^{\left(i\right)}\right)^{2}.
\end{equation}
Fig. \ref{fig:non_mono} (c) and (g) shows that the convergence time
of $\delta\eta$ grows with $\alpha$. Non-monotonic responses are
more difficult to train, since they have a larger convergence time.
At large times $\delta\eta$ decays approximately as a power-law,
roughly scaling as $\tau^{\approx-0.5}$ for the allostery inspired
response, and a quicker $\tau^{\approx-1.0}$ decay for the uniaxial
strain.

Since training reduces the energy along the training paths, we also
measure the elastic energy along the path as a function of the number
of training cycles. Surprisingly, the $\alpha$ dependent time scale
is not reflected in the relaxation of the energy. Fig. \ref{fig:non_mono}
(d) and (h) shows that the energy approximately scales as $\tau^{-1}$
independent of $\alpha$. This appears to be a robust feature common
to many of the examples considered. Below we present an argument for
that scaling. 

\subsection{XOR gate via bifurcations}
Within linear response applying two different strains results in a response that is the sum of the two individual responses. In the nonlinear regime this restriction does not apply. We consider logic gates with two inputs and a single output as an example as a non-additive mapping. Here we aim at realizing the XOR gate (AND and OR can be realized using similar ideas). Logic gates were previously achieved by design of 3d-printed structures\citep{jiang2019bifurcation,zanaty2020reconfigurable}. 

The XOR gate is defined in Fig. \ref{fig:XOR}(a) has two discrete
inputs and a single output. Note that negating the two inputs does
not negate the output. We assign two sources and a single target,
each to a randomly chosen pair of adjacent nodes (as in allostery).
The corresponding strain are $\eta_{S1},\eta_{S2}$ for the sources
and $\eta_{T}$ for the target. The four discrete state correspond
to the maximal strain amplitude, where $\left|\eta_{S1}\right|=\left|\eta_{S2}\right|=1$.
Our goal is to train the system so that actuating the two source sites to one of the four states  will yield  $\eta_{T}=\pm1$ as defined by the XOR gate. To train
these responses we interpolate the discrete states into continuous
paths that bifurcate from the unstrained state at the origin. The training
paths, shown in Fig. \ref{fig:XOR}(b), are composed of two curves:
I) $\eta_{T}=-\eta_{S1}^{2}=-\eta_{S2}^{2}$ with  $\eta_{S1}=\eta_{S2}$,
and II) $\eta_{T}=\eta_{S1}^{2}=\eta_{S2}^{2}$ with $\eta_{S1}=-\eta_{S2}$.
Note these curves only meet at the unstrained state ($\eta_{S1}=\eta_{S2}=\eta_{T}=0$)
and depending on the applied strains the system follows one of the four branches. We note that multi-branched responses have also been studied in Ref. \citep{pinson2017self,lubbers2019excess}.  

Fig. \ref{fig:XOR}(b) shows the measured response after a large number of training cycles.
As can be seen, it reasonably matches the training paths, denoted by the
dashed curves. Fig. \ref{fig:XOR}(c) shows $\delta\eta$ as a function
of cycles; also here it decays as approximately as $\tau^{-0.5}$.
The energy vs. number of cycles is shown in \ref{fig:XOR}(d) approximately
follows the $\tau^{-1}$ scaling. In this example we also test the
effect of varying the aging rate, $\Gamma=\gamma k$, but find little
change to the functional behavior. 

\subsection{Sequence dependent responses}
\begin{figure}
\includegraphics[scale=0.5]{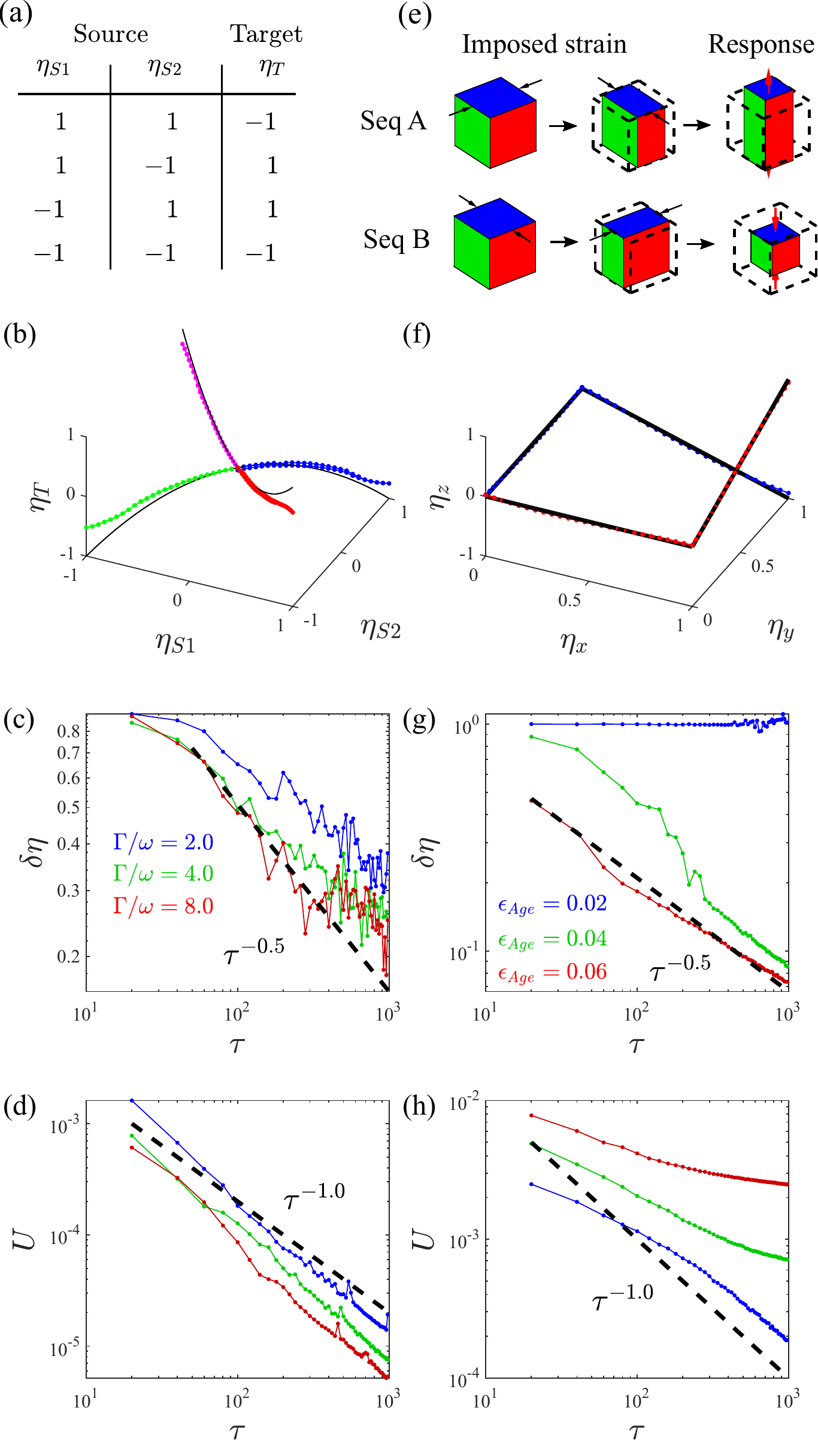}

\caption{Left: Training the XOR logic gate define in (a) as an example of a
non-additive response. Right: An example of a sequence dependent material,
where the order of compressions along the x and y axis determine the
response along the z-axis (e). (b) and (f) the corresponding training
paths. (c) and (g) the convergence of the response as a function of
the number of training cycles. (d) and (h) the energy along the training
path as a function of the number of training cycles. For the XOR gate
we show that varying the aging rate $\Gamma=\gamma k$ with respect
to training frequency $\omega$ has little effect. 
For the sequence dependent material small strains are less successful. Simulation parameters for XOR gate:  $\epsilon_{Age}=0.2$, average is over 20 realizations. For sequence dependent material $\Gamma=2.0$ and 50 realizations. \label{fig:XOR} }
\end{figure}

 We have shown that bifurcations allow the system to traverse different branches as a function of the two (or more) imposed strains. Here we consider a bifurcation where the branch is selected based on the order of applied strains, rather than their value. We consider a three dimensional system with two sources, defined
to be the uniaxial deformation along the x and y direction, while
target deformation is along the z-axis. The corresponding normalized
strain along these directions are denoted by $\eta_{x},\eta_{y}$
and $\eta_{z}$. The goal of training is shown in Fig. \ref{fig:XOR}(e):
compressing along the x-axis and then the y-axis results in an expansion
along the z-axis. Reversing the order of applied strains yields compression
along the z-axis. 

Such a response is realized through two training paths that bifurcate from the unstrained state, as shown in Fig. \ref{fig:XOR}(f). Each sequence is composed of two linear segments: Seq A: $\left(\eta_{x},\eta_{y},\eta_{z}\right)\rightarrow\left(0,0,0\right)\rightarrow\left(1,0,0\right)\rightarrow\left(1,1,1\right)$.
Seq B: $\left(\eta_{x},\eta_{y},\eta_{z}\right)\rightarrow\left(0,0,0\right)\rightarrow\left(0,1,0\right)\rightarrow\left(1,1,-1\right)$.
Because of the third dimension the two paths do not cross. To train such a path we alternate between training each of the two branches. 

The measured response, $\eta_{z}$ as a function of $\eta_{x}$ and
$\eta_{y}$ is shown in Fig. \ref{fig:XOR}(g) along with the training
path (dashed curve). Training is successful, despite the sharp corner
of the training path. Fig. \ref{fig:XOR}(h) shows the convergence
of the target response, $\delta\eta$ as a function of the number
of the training cycles. Here, we consider the effect of varying the
strain amplitude, $\epsilon_{Age}$ in the range of $2\%$ to $6\%$.
At the largest strain amplitude $\epsilon_{Age}$ training is highly
successful and there is a clear sequence dependence. At small strains
training is unsuccessful, suggesting that large strains are needed
to access the non-linear regime.

We also plot the energy as a function of the number of cycles for
the different strain amplitudes, $\epsilon_{Age}$. Here, the functional
dependence depends on $\epsilon_{Age}$: while for small $\epsilon_{Age}$
it approximately scales as $\tau^{-1.0}$ , at large strains it decays
more slowly. Attaining this response at large strains is therefore
more difficult, and since it is the only example that deviates from
that convergence it could be an indicator of a new regime. 

\subsection{Evolution of the elastic energy}
Over a broad range of parameters and
training paths the energy along the training path  decays approximately as $\tau^{-1}$. We explain this trend through a simple argument. The change in rest length
over a cycle scales as the force, or the energy divided by the strain:
$\Delta\ell_{0}\sim f\Delta\tau\propto U/\epsilon_{Age}\Delta\tau$. Assuming small deformations, the change in energy over a cycle is
given by: 
\begin{align}
\Delta U & =\frac{\partial U}{\partial\ell_{0}}\Delta\ell_{0}\\
 & \sim-\frac{1}{a\epsilon_{Age}}U^{2}\Delta\tau,
\end{align}
where we estimated the derivative by, $\frac{\partial U}{\partial\ell_{0}}\sim\frac{U}{a}$,
and $a$ is a microscopic length scale. Solving the differential equation
yields $U\propto\epsilon_{Age}\tau^{-1}$ as observed. This also predicts
the prefactor $\epsilon_{Age}$, which is consistent with simulations. 

\section{Conclusions}
We have demonstrated that compound spring-dashpot
bonded networks can be trained along highly nonlinear elaborate paths
that curve and bifurcate. Our results are both valid for global responses
as well as bond specific responses. These allow functionality that
is usually not easily achieved by design. The nonmonotonic paths could
possibly be used to transform the frequency of a signal. The bifurcating
paths can be programmed to function as logic gates and allows materials
whose response depends on the sequence of the imposed strains.

Though training is successful, the slowness of the dynamics hints
at a difficulty. The convergence rate for the nonmonotonic responses
becomes increasing slow as the path increasingly curves. The slow
convergence could possibly be due to the unintentional formation of
additional competing low energy modes\citep{stern2017complexity,tachi2017self,dieleman2020jigsaw}.
The desired response is then only achieved when the energy along the
target path is made smaller than the remaining competing directions.
The density of states provides a rudimentary measurement of these
low energy modes. In the Appendix B we show that indeed
increasing $\alpha$ yields more low energy modes. Characterizing the entire energy landscape, however, is a difficult task. A different source
of difficultly is seen in the slower than usual decay of the energy
in the sequence dependent responses. This could point to a new regime,
where the system cannot attain the prescribed low energy mode, possibly an indication of a limit to the system's capacity. 

\section*{Acknowledgements}
We would like to thank Dov Levine, Andrea
J. Liu, Sidney R. Nagel, Nidhi Pashine and Menachem Stern for enlightening
discussions. We acknowledge the University of Chicago Research Computing
Center for support of the initial steps of this work. We are grateful
to the ATLAS cluster at the Technion for providing computing resources.
This work was  supported by the Israel Science Foundation (grant 2385/20).

\section*{Appendix A: numerical methods}
\subsection*{Packing preparation}
We prepare our networks from packings
of soft spheres with repulsive harmonic interactions\citep{Ohern,Goodrich2013}. For every pair
of overlapping particles (the distance $\left|r_{i}-r_{j}\right|$
falls below the sum of the radii $R_{i}+R_{j}$) the energy is given
by:
\[
U_{ij}=\frac{k}{2}\sum\left(1-\frac{\left|r_{i}-r_{j}\right|}{R_{i}+R_{j}}\right)^{2}.
\]
To attain force balance we minimize the energy using the FIRE algorithm
\citep{FIRE}, under a constant imposed pressure. We then construct
networks attaching the centers of overlapping particles with springs.
The packings can then be discarded. For simplicity we set the spring
constants to be unity and the initial rest lengths to be the interparticle
distance, so that the initial networks are unstressed. 

\subsection*{Details of networks}
With the exception of the sequential
materials we use two dimensional packing with approximately 500 nodes.
Two dimensional systems allow us to study signaling (allostery) over
long distances with relatively fewer nodes than three dimensions.
The boundary conditions, for simplicity, are chosen to be periodic.
For the nonmonotonic allostery responses and the XOR gate the excess
coordination number $\Delta Z=\frac{2N_{B}}{N}-2d\approx0.05$. Here,
$N_{b}$ is the number of bonds, $N$ is the number of nodes and $d$
is the dimensionality of the system. These spatially varying responses
require a small $\Delta Z$, where elasticity is anomalously long
ranged\citep{ellenbroek2006critical,lerner2014breakdown}. For the
nonmonotonic global responses we use $\Delta Z\approx0.15$ which
is not particularly small. The networks for the sequential response
are three dimensional with $\Delta Z\approx0.4$. Unless indicated
otherwise, we average over approximately 50 realizations.

\section*{Training protocol} 
Following Ref. \citep{hexner2019periodic}
we discritize the strain path to small steps. Every step consists
of updating the strain, minimizing the elastic energy to attain force
balance and then updating the rest lengths of springs. As discussed
in the main text, the change in rest length is proportional to the
stress on the bond: 
\[
\Delta\ell_{i,0}=\gamma k\left(\ell_{i}-\ell_{i,0}\right)\Delta t.
\]
When the step size is made small then then the dynamics may be considered
quasistatic. We verify, that finer discretization does not alter our
results. 

\section*{Appendix B: Density of states}

\begin{figure}[H]

\includegraphics[scale=0.6]{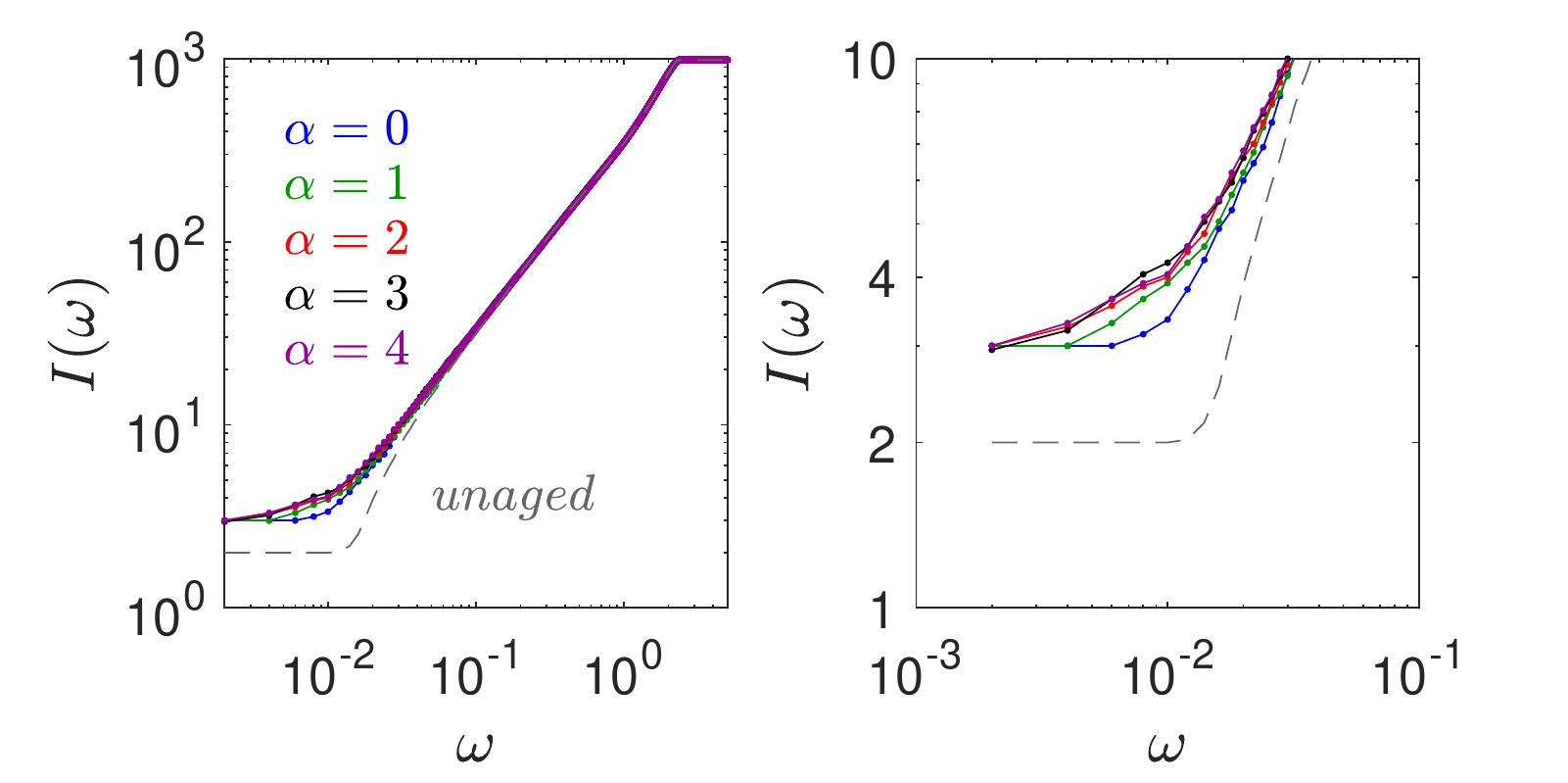}

\caption{The integrated density of states. Left: over the entire frequency
range. Right: zoom in at small frequencies. Larger $\alpha$ have
an increased number of low energy modes in comparison to the small
$\alpha$ curves.\label{fig:Iw}}

\end{figure}

Convergence of training in the nonmonotonic
responses exhibits a growing time scale as $\alpha$ is increased.
One possible explanation for this is that low energy modes are created
during training that compete with the trained path. To explore this
possibility we consider the normal modes around the origin. The density
of states $D\left(\omega\right)$ is defined as the number of modes
in the range $\left[\omega,\omega+d\omega\right]$ divided by $d\omega$
and the number of particles. We focus on the integrated density of
states $I\left(\omega\right)=N\int_{0}^{\omega}D\left(\omega'\right)d\omega'$
shown in Fig. \ref{fig:Iw} since it allows us to count the number
of low energy modes. For the untrained two dimensional system there
are precisely two transnational zero modes due to the periodic conditions,
and therefore $I\left(\omega\right)$ has a plateau at small frequencies.
The trained systems have an additional near zero mode corresponding
to the trained response, and therefore in that regime $I\left(\omega\right)\approx3$.
The trained configurations have an increased number of low energy modes
in comparison to the untrained systems. Furthermore, increasing $\alpha$
also increases the number of low energy modes. This trend is most
evident for $\alpha=0,1,2$ though $\alpha=3,4$ are very close to
$\alpha=2.$ While this is suggestive of this competing low energy
mode picture, the density of states only characterizes infinitesimal
deformations. 

\bibliographystyle{plain}
\bibliography{biblo}

\begin{thebibliography}{10}

\bibitem{bar2020geometric}
Yohai Bar-Sinai, Gabriele Librandi, Katia Bertoldi, and Michael Moshe.
\newblock Geometric charges and nonlinear elasticity of two-dimensional elastic
  metamaterials.
\newblock {\em Proceedings of the National Academy of Sciences},
  117(19):10195--10202, 2020.

\bibitem{bertoldi2010negative}
Katia Bertoldi, Pedro~M Reis, Stephen Willshaw, and Tom Mullin.
\newblock Negative poisson's ratio behavior induced by an elastic instability.
\newblock {\em Advanced materials}, 22(3):361--366, 2010.

\bibitem{FIRE}
Erik Bitzek, Pekka Koskinen, Franz G\"ahler, Michael Moseler, and Peter
  Gumbsch.
\newblock Structural relaxation made simple.
\newblock {\em Phys. Rev. Lett.}, 97:170201, 2006.

\bibitem{chen2014nonlinear}
Bryan Gin-ge Chen, Nitin Upadhyaya, and Vincenzo Vitelli.
\newblock Nonlinear conduction via solitons in a topological mechanical
  insulator.
\newblock {\em Proceedings of the National Academy of Sciences},
  111(36):13004--13009, 2014.

\bibitem{coulais2016combinatorial}
Corentin Coulais, Eial Teomy, Koen de~Reus, Yair Shokef, and Martin van Hecke.
\newblock Combinatorial design of textured mechanical metamaterials.
\newblock {\em Nature}, 535(7613):529, 2016.

\bibitem{dieleman2020jigsaw}
Peter Dieleman, Niek Vasmel, Scott Waitukaitis, and Martin van Hecke.
\newblock Jigsaw puzzle design of pluripotent origami.
\newblock {\em Nature Physics}, 16(1):63--68, 2020.

\bibitem{ellenbroek2006critical}
Wouter~G Ellenbroek, Ell{\'a}k Somfai, Martin van Hecke, and Wim van Saarloos.
\newblock Critical scaling in linear response of frictionless granular packings
  near jamming.
\newblock {\em Physical review letters}, 97(25):258001, 2006.

\bibitem{Goodrich2013}
Carl~P. Goodrich, Wouter~G. Ellenbroek, and Andrea~J. Liu.
\newblock Stability of jammed packings i: the rigidity length scale.
\newblock {\em Soft Matter}, 9:10993--10999, 2013.

\bibitem{goodrich2015principle}
Carl~P Goodrich, Andrea~J Liu, and Sidney~R Nagel.
\newblock The principle of independent bond-level response: Tuning by pruning
  to exploit disorder for global behavior.
\newblock {\em Physical review letters}, 114(22):225501, 2015.

\bibitem{grima2000auxetic}
Joseph~N Grima and Kenneth~E Evans.
\newblock Auxetic behavior from rotating squares.
\newblock {\em Journal of Materials Science Letters}, 19:1563, 2000.

\bibitem{hexner2018role}
Daniel Hexner, Andrea~J Liu, and Sidney~R Nagel.
\newblock Role of local response in manipulating the elastic properties of
  disordered solids by bond removal.
\newblock {\em Soft matter}, 14(2):312--318, 2018.

\bibitem{hexner2019periodic}
Daniel Hexner, Andrea~J. Liu, and Sidney~R. Nagel.
\newblock Periodic training of creeping solids.
\newblock {\em Proceedings of the National Academy of Sciences}, 2020.

\bibitem{hexner2020effect}
Daniel Hexner, Nidhi Pashine, Andrea~J Liu, and Sidney~R Nagel.
\newblock Effect of directed aging on nonlinear elasticity and memory formation
  in a material.
\newblock {\em Physical Review Research}, 2(4):043231, 2020.

\bibitem{jiang2019bifurcation}
Yijie Jiang, Lucia~M Korpas, and Jordan~R Raney.
\newblock Bifurcation-based embodied logic and autonomous actuation.
\newblock {\em Nature communications}, 10(1):1--10, 2019.

\bibitem{kim2019conformational}
Jason~Z Kim, Zhixin Lu, Steven~H Strogatz, and Danielle~S Bassett.
\newblock Conformational control of mechanical networks.
\newblock {\em Nature Physics}, 15(7):714--720, 2019.

\bibitem{lakes_r_Science_1987}
Roderic Lakes.
\newblock Foam structures with a negative poisson{\textquoteright}s ratio.
\newblock {\em Science}, 235(4792):1038--1040, 1987.

\bibitem{lerner2014breakdown}
Edan Lerner, Eric DeGiuli, Gustavo D{\"u}ring, and Matthieu Wyart.
\newblock Breakdown of continuum elasticity in amorphous solids.
\newblock {\em Soft Matter}, 10(28):5085--5092, 2014.

\bibitem{lubbers2019excess}
Luuk~A Lubbers and Martin van Hecke.
\newblock Excess floppy modes and multibranched mechanisms in metamaterials
  with symmetries.
\newblock {\em Physical Review E}, 100(2):021001, 2019.

\bibitem{maxwell1867iv}
James~Clerk Maxwell.
\newblock Iv. on the dynamical theory of gases.
\newblock {\em Philosophical transactions of the Royal Society of London},
  (157):49--88, 1867.

\bibitem{meeussen2020topological}
Anne~S Meeussen, Erdal~C O{\u{g}}uz, Yair Shokef, and Martin van Hecke.
\newblock Topological defects produce exotic mechanics in complex
  metamaterials.
\newblock {\em Nature Physics}, pages 1--5, 2020.

\bibitem{mitchell2016strain}
Michael~R Mitchell, Tsvi Tlusty, and Stanislas Leibler.
\newblock Strain analysis of protein structures and low dimensionality of
  mechanical allosteric couplings.
\newblock {\em Proceedings of the National Academy of Sciences},
  113(40):E5847--E5855, 2016.

\bibitem{miura1985method}
Koryo Miura.
\newblock Method of packaging and deployment of large membranes in space.
\newblock {\em Title The Institute of Space and Astronautical Science Report},
  618:1, 1985.

\bibitem{Ohern}
Corey~S. O'Hern, Leonardo~E. Silbert, Andrea~J. Liu, and Sidney~R. Nagel.
\newblock Jamming at zero temperature and zero applied stress: The epitome of
  disorder.
\newblock {\em Phys. Rev. E}, 68:011306, 2003.

\bibitem{pashine2019directed}
Nidhi Pashine, Daniel Hexner, Andrea~J Liu, and Sidney~R Nagel.
\newblock Directed aging, memory, and nature's greed.
\newblock {\em Science advances}, 5(12):eaax4215, 2019.

\bibitem{pinson2017self}
Matthew~B Pinson, Menachem Stern, Alexandra~Carruthers Ferrero, Thomas~A
  Witten, Elizabeth Chen, and Arvind Murugan.
\newblock Self-folding origami at any energy scale.
\newblock {\em Nature communications}, 8(1):1--8, 2017.

\bibitem{reid2018auxetic}
Daniel~R Reid, Nidhi Pashine, Justin~M Wozniak, Heinrich~M Jaeger, Andrea~J
  Liu, Sidney~R Nagel, and Juan~J de~Pablo.
\newblock Auxetic metamaterials from disordered networks.
\newblock {\em Proceedings of the National Academy of Sciences},
  115(7):E1384--E1390, 2018.

\bibitem{rocks2017designing}
Jason~W Rocks, Nidhi Pashine, Irmgard Bischofberger, Carl~P Goodrich, Andrea~J
  Liu, and Sidney~R Nagel.
\newblock Designing allostery-inspired response in mechanical networks.
\newblock {\em Proceedings of the National Academy of Sciences},
  114(10):2520--2525, 2017.

\bibitem{rocks2019limits}
Jason~W Rocks, Henrik Ronellenfitsch, Andrea~J Liu, Sidney~R Nagel, and Eleni
  Katifori.
\newblock Limits of multifunctionality in tunable networks.
\newblock {\em Proceedings of the National Academy of Sciences},
  116(7):2506--2511, 2019.

\bibitem{stern2017complexity}
Menachem Stern, Matthew~B Pinson, and Arvind Murugan.
\newblock The complexity of folding self-folding origami.
\newblock {\em Physical Review X}, 7(4):041070, 2017.

\bibitem{tachi2017self}
Tomohiro Tachi and Thomas~C Hull.
\newblock Self-foldability of rigid origami.
\newblock {\em Journal of Mechanisms and Robotics}, 9(2), 2017.

\bibitem{yan2017architecture}
Le~Yan, Riccardo Ravasio, Carolina Brito, and Matthieu Wyart.
\newblock Architecture and coevolution of allosteric materials.
\newblock {\em Proceedings of the National Academy of Sciences},
  114(10):2526--2531, 2017.

\bibitem{zanaty2020reconfigurable}
Mohamed Zanaty, Hubert Schneegans, Ilan Vardi, and Simon Henein.
\newblock {Reconfigurable Logic Gates Based on Programable Multistable
  Mechanisms}.
\newblock {\em Journal of Mechanisms and Robotics}, 12(2), 02 2020.
\newblock 021111.

\end{thebibliography}

\end{document}